\documentclass[runningheads,dvipsnames]{llncs}

\usepackage[setpagesize=false,hidelinks]{hyperref}
\usepackage{listings}
\usepackage{fixmetodonotes}
\usepackage{tikz}
\usetikzlibrary{positioning,arrows,calc}
\usepackage{pgfplots}
\usepackage{amsmath}
\usepackage{fancyvrb}
\usepackage[]{xcolor}

\title{Declarative Programming for Microcontrollers - Datalog on Arduino}
\titlerunning{Datalog on Arduino}

\author{Mario Wenzel \and Stefan Brass}
\institute{Martin-Luther-Universit\"at Halle-Wittenberg,
	Institut f\"ur Informatik,\\
	Von-Seckendorff-Platz~1, D-06099 Halle (Saale), Germany\\
	\email{mario.wenzel@informatik.uni-halle.de},
	\email{brass@informatik.uni-halle.de}}

\makeatletter

\def\define#1{\@ifnextchar [{\@MYargdef#1}{\@MYargdef#1[0]}}

\def\@MYargdef#1[#2]#3{\@ifdefinable #1{\@MYreargdef#1[#2]{#3}}}

\def\redefine#1{\edef\@tempa{\expandafter\@cdr\string
  #1\@nil}\@ifundefined{\@tempa}{\@latexerr{\string#1\space undefined}\@ehc
    }{}\@ifnextchar [{\@MYreargdef#1}{\@MYreargdef#1[0]}}

\catcode`\?=11\relax
\def\@MYreargdef#1[#2]#3{\@tempcnta#2\relax\let#1\relax
\edef\@tempa{\def#1}\@tempcntb \@ne
\let\@?@?\relax\@whilenum\@tempcnta>0
\do{\edef\@tempa{\@tempa\@?@?\the\@tempcntb}\advance\@tempcntb \@ne \advance
\@tempcnta \m@ne}\let\@?@?##\@tempa{#3}}
\catcode`\?=12\relax

\makeatother

\renewcommand{\bmod}%
{\mskip-\medmuskip \mkern5mu \mathbin{\idFont mod} \mkern5mu \mskip-\medmuskip}

\define{\mathsym}[1]{\relax\ifmmode#1\else
	\errmessage{Mathematical symbol outside math mode}\fi}
\define{\idFont}{\sf}
\define{\id}[1]{\mathsym{{\idFont #1}}}

\define{\m}[1]{$#1$}
\define{\M}[1]{$$#1$$}

\define{\gt}{\mathsym{\mathchar 12606\relax}} 
\define{\lt}{\mathsym{\mathchar 12604\relax}} 

\define{\paren}[1]{(#1)}
\define{\parenA}[1]{(#1)}
\define{\parenB}[1]{\bigl(#1\bigr)}
\define{\parenC}[1]{\Bigl(#1\Bigr)}
\define{\parenAuto}[1]{\left(#1\right)}
\define{\set}[1]{\lbrace#1\rbrace}
\define{\setA}[1]{\lbrace#1\rbrace}
\define{\setB}[1]{\bigl\lbrace#1\bigr\rbrace}
\define{\setC}[1]{\Bigl\lbrace#1\Bigr\rbrace}
\define{\setCond}[2]{\lbrace#1\mathrel{\vert}#2\rbrace}
\define{\setCondA}[2]{\lbrace#1\mathrel{\vert}#2\rbrace}
\define{\setCondB}[2]{\bigl\lbrace#1\bigm\vert#2\bigr\rbrace}
\define{\setCondC}[2]{\Bigl\lbrace#1\Bigm\vert#2\Bigr\rbrace}

\define{\union}{\cup}
\define{\unionUntil}{\union\cdots\union}
\define{\unionMulti}[2]{\bigcup_{#1}^{#2}}
\define{\intersect}{\cap}
\define{\intersectUntil}{\intersect\cdots\intersect}
\define{\intersectMulti}[2]{\bigcap_{#1}^{#2}}
\define{\timesUntil}{\times\cdots\times}
\define{\setsize}[1]{{\left\vert #1 \right\vert}}
\define{\compl}[1]{\overline{#1}}

\define{\powerset}[1]{2^{#1}}
\define{\inB}{\;\in\;}

\define{\defEq}{:=}
\define{\defEqB}{\;:=\;}
\define{\defIff}{\;\mathsym{:\Longleftrightarrow}\;}
\define{\metaThen}{\mathsym{\;\Longrightarrow\;}}
\define{\metaIf}{\mathsym{\;\Longleftarrow\;}}
\define{\metaIff}{\mathsym{\;\Longleftrightarrow\;}}

\define{\lneg}{\mathsym{\neg}}
\define{\lif}{\mathsym{\leftarrow}}
\define{\lifDup}{\mathsym{\Leftarrow}}
\define{\lthen}{\mathsym{\rightarrow}}
\define{\liff}{\mathsym{\leftrightarrow}}
\define{\lfalse}{\id{false}}
\define{\ltrue}{\id{true}}

\define{\lorUntil}{\lor\cdots\lor}
\define{\landUntil}{\land\cdots\land}
\define{\lorMulti}[2]{\bigvee_{#1}^{#2}}
\define{\landMulti}[2]{\bigwedge_{#1}^{#2}}

\define{\lfor}[4]{#1\:#2\:#3:#4}
\define{\lexistsQ}{\exists}
\define{\lexists}[3]{\lfor{\lexistsQ}{#2}{#1}{#3}}
\define{\lallQ}{\forall}
\define{\lall}[3]{\lfor{\lallQ}{#2}{#1}{#3}}

\define{\answerPred}{\id{answer}}

\define{\natNum}{\mathsym{{\rm l\kern-0.13em N}}}

\define{\realNum}{\mathsym{{\rm I\kern-0.14em R}}}

\define{\struct}[1]{\langle#1\rangle}
\define{\twoCases}[4]{\left\{\begin{array}{l@{\kern10pt}l}
				#1&\mbox{#2}\\#3&\mbox{#4}
				\end{array}\right.}
\define{\until}{, \ldots,}
\define{\seqOf}[1]{#1^*}
\define{\emptySeq}{\epsilon}
\define{\w}{\id{w}}

\define{\code}[1]{{\tt #1}} 

{\begin{center}\begin{tt}\begin{tabular}{@{}l@{}}}%
{\end{tabular}\end{tt}\end{center}}

	{\begin{center}\begin{tt}\begin{tabular}{@{}l@{ }l@{}}}%
	{\end{tabular}\end{tt}\end{center}}

\define{\U}{{\char95}} 
\define{\LT}{{\char60}} 
\define{\GT}{{\char62}} 
\define{\B}{{\char92}} 
\define{\AMP}{{\char38}} 
\define{\D}{{\char36}} 
\define{\SN}{{\char126}} 
\define{\HASH}{{\char35}} 
\define{\Q}{{\char34}} 
\define{\PCT}{{\char37}} 
\define{\LB}{{\char123}} 
\define{\RB}{{\char125}} 
\define{\HAT}{{\char94}} 

\define{\ALPH}{\id{ALP\kern-0.08em H}}
\define{\LOG}{\id{LOG}}
\define{\VARS}{\id{V\kern-0.17em ARS}}
\define{\var}{\mathsym{X}}
\define{\varA}{\mathsym{X}}
\define{\varB}{\mathsym{Y}}
\define{\varC}{\mathsym{Z}}
\define{\const}{\mathsym{c}}
\define{\constA}{\mathsym{a}}
\define{\constB}{\mathsym{b}}
\define{\constC}{\mathsym{c}}
\define{\constD}{\mathsym{d}}
\define{\data}{\mathsym{d}}
\define{\dataSet}{\mathsym{{\cal D}}}
\define{\sig}{\Sigma}
\define{\SORTS}{\id{{\cal S}}}
\define{\sort}{\id{s}}
\define{\PREDS}{\id{{\cal P}}}
\define{\pred}{\mathsym{p}}
\define{\predA}{\mathsym{p}}
\define{\predB}{\mathsym{q}}
\define{\predC}{\mathsym{r}}
\define{\predD}{\mathsym{s}}
\define{\arity}{\mathsym{n}}
\define{\level}{l}
\define{\FUNS}{\id{{\cal F}}}
\define{\fun}{\id{f}}
\define{\args}{\alpha}
\define{\argsB}{\beta}
\define{\argSorts}{\alpha}
\define{\resSort}{\rho}
\define{\interp}{\id{{\cal I}}}
\define{\interpB}{\id{{\cal J}}}
\define{\ass}{\id{{\cal A}}}
\define{\iV}[2]{(#1,#2)}
\define{\eval}[2]{#1\lbrack\kern-0.15em\lbrack#2\rbrack\kern-0.15em\rbrack}
\define{\evalV}[3]{\eval{\iV{#1}{#2}}{#3}}
\define{\varDecl}{\nu}
\define{\TERMS}{\id{T\kern-0.1em E}}
\define{\term}{\mathsym{t}}
\define{\termB}{\mathsym{u}}
\define{\termC}{\mathsym{v}}
\define{\argA}{\id{a}}
\define{\argB}{\id{b}}
\define{\argC}{\id{c}}
\define{\AT}{\id{AT}}
\define{\FO}{\id{FO}}
\define{\fo}{\varphi}
\define{\foB}{\psi}
\define{\fos}{\Phi}
\define{\modify}[3]{#1\langle#2/#3\rangle}
\define{\impl}{\vdash}
\define{\subst}{\theta}
\define{\substB}{\sigma}
\define{\mgu}{\id{mgu}}
\define{\doSubst}[2]{#2#1}
\define{\doSubstB}[2]{(#2)\,#1}
\define{\HU}{\id{{\cal U}}}
\define{\HB}{\id{{\cal B}}}
\define{\HSet}{\id{H}}
\define{\ground}{\id{ground}}
\define{\LAT}{\id{{\cal M}}}
\define{\LATB}{\id{{\cal N}}}

\define{\emptyClause}{{\hbox{%
	\setlength{\unitlength}{0.24ex}%
	\begin{picture}(5,5)(0,0)
	\put(0,0){\line(0,1){5}}
	\put(0,0){\line(1,0){5}}
	\put(0,5){\line(1,0){5}}
	\put(5,0){\line(0,1){5}}
	\end{picture}}}}

\define{\lit}{\mathsym{L}}
\define{\litA}{\mathsym{A}}
\define{\litB}{\mathsym{B}}
\define{\litC}{\mathsym{C}}
\define{\Body}{\mathsym{{\cal B}}}
\define{\F}{\id{F}} 
\define{\ruleFun}[1]{\id{r}_{#1}}

\define{\prog}{\mathsym{P}}
\define{\T}[1]{\mathsym{{\idFont T}}_{#1}}
\define{\TP}{\T{\prog}}
\define{\Tneg}[2]{\mathsym{{\idFont T}}_{#1,#2}}
\define{\lub}{\id{lub}}
\define{\glb}{\id{glb}}
\define{\lfp}{\id{l\kern-0.1em f\kern-0.1em p}}
\define{\db}{\mathsym{D}}
\define{\minMod}{\mathsym{{\cal M}}}

\define{\IDBPred}{\id{IDB}}

\define{\I}{\id{{\cal I}}}
\define{\J}{\id{{\cal J}}}

\define{\nf}{\mathop{\id{not\kern0.2em}}}
\define{\nfPred}[1]{\mathop{\id{not{\U}}#1}}

\define{\free}{\id{f}}
\define{\bound}{\id{b}}

\define{\bp}{\id{bp}}
\define{\binding}{\beta}
\define{\bindingSet}{{\cal B}}

\define{\vars}{\mathsym{{\cal X}}}
\define{\varsA}{\mathsym{{\cal X}}}
\define{\varsB}{\mathsym{{\cal Y}}}
\define{\varsC}{\mathsym{{\cal Z}}}

\define{\inputVars}{\id{input}}

\define{\freeVar}{\id{vars}}
\define{\varsOf}{\id{vars}}
\define{\boundPos}{\id{bound}^+}
\define{\boundNeg}{\id{bound}^-}
\define{\unboundPos}{\id{unbound}^+}
\define{\unboundNeg}{\id{unbound}^-}
\define{\err}{\id{err}}

\newcounter{ProgramLine}
\newcounter{FirstLine}

\newenvironment{progTabular}{%
	\addtocounter{ProgramLine}{-1}%
	\setcounter{FirstLine}{1}%
	\ifvmode\vspace{5mm}\fi%
	\begin{list}{\(\bullet\)\hfill}{
		\parskip 3pt plus 1pt
		\labelwidth 0pt
		\labelsep 0pt
		\leftmargin 8mm
		\listparindent \parindent
		\topsep 4mm
		\parsep 3pt plus 1pt
		\itemsep 0pt
		\partopsep 0pt
		\itemindent 0pt
		\rightmargin 0pt
	}
	\item[]
	\begin{tabular}{@{}r@{\hspace{2mm}}l@{}}}%
	{\end{tabular}%
		\end{list}}

\newenvironment{progPart}[1]{%
		\setcounter{ProgramLine}{#1}%
		\begin{progTabular}}%
	{\end{progTabular}}

	{\end{progTabular}%
		\end{tt}}

	{\end{progTabular}%
		\end{tt}}

\define{\tabX}[1]{\ifnum\value{FirstLine}=1\setcounter{FirstLine}{0}\else
	\ifhmode\\[0.5pt]\fi\fi
	\stepcounter{ProgramLine}(\arabic{ProgramLine})&
	\hspace*{#1}}
\define{\tabA}{\tabX{0cm}}
\define{\tabB}{\tabX{8mm}}
\define{\tabC}{\tabX{16mm}}
\define{\tabD}{\tabX{24mm}}
\define{\tabE}{\tabX{32mm}}
\define{\tabF}{\tabX{40mm}}
\define{\tabG}{\tabX{50mm}}
\define{\tabH}{\tabX{60mm}}
\define{\tabBox}[1]{\vrule height0pt depth0pt width0pt\hbox to14mm{#1\ \hfil}}

\define{\IF}{{\bf if }}
\define{\THEN}{{\bf then }}
\define{\ELSE}{{\bf else }}
\define{\FI}{{\bf fi}}
\define{\FOREACH}{{\bf foreach }}
\define{\FOR}{{\bf for }}
\define{\TO}{{\bf to }}
\define{\FROM}{{\bf from }}
\define{\IN}{{\bf in }}
\define{\WITH}{{\bf with }}
\define{\AND}{{\bf and }}
\define{\NOT}{{\bf not }}
\define{\TRUE}{{\bf true}}
\define{\FALSE}{{\bf false}}
\define{\DO}{{\bf do }}
\define{\OD}{{\bf od}}
\define{\WHILE}{{\bf while }}
\define{\BREAK}{{\bf break}}
\define{\PROCEDURE}{{\bf procedure }}
\define{\BEGIN}{{\bf \code{\LB}} }
\define{\END}{{\bf \code{\RB}} }
\define{\RETURN}{{\bf return }}
\define{\LET}{{\bf let }}
\define{\NEW}{{\bf new }}
\define{\COMPUTE}{{\bf compute }}
\define{\APPEND}{{\bf append }}
\define{\PRINT}{{\bf print }}
\define{\BOOL}{{\bf bool }}
\define{\NIL}{{\bf nil }}
\define{\OUTPUT}{{\bf output }}
\define{\INSERT}{{\bf insert }}
\define{\INTO}{{\bf into }}
\define{\CHOOSE}{{\bf choose }}

\define{\COMMENT}[1]{{\rm /\raisebox{-.6ex}{*} #1 \raisebox{-.6ex}{*}/}}

\define{\uri}[1]{\href{#1}{\code{[#1]}}}
\define{\uriDiff}[2]{\href{#1}{\code{[#2]}}}
\define{\uriPlain}[1]{\href{#1}{\code{#1}}}
\define{\uriDiffPlain}[2]{\href{#1}{\code{#2}}}

\define{\CPP}{C{\tt ++}}

\define{\activePart}[1]{\id{a}(#1)}
\define{\delayedPart}[1]{\id{d}(#1)}

\define{\lineno}[1]{\hbox to 1.6em{\hfil\m{\lbrack#1\rbrack}}}
\define{\lineref}[1]{\m{\lbrack#1\rbrack}}
\define{\comment}[1]{\mbox{// #1}}

\define{\C}{\mathsym{C}}

\define{\ruleNo}{\rho}

\define{\edbPredA}{\id{e}}
\define{\edbPredB}{\id{r}}
\define{\edbPredC}{\id{s}}

\define{\state}{\mathsym{{\cal S}}}
\define{\Rule}{\mathsym{{R}}}

\define{\vertices}{\mathsym{{\cal V}}}
\define{\edges}{\mathsym{{\cal E}}}

\define{\called}[2]{(\lineref{#1},\lineref{#2})}

\define{\factSeq}{\mathsym{{\cal S}}}
\define{\fact}{\mathsym{F}}

\define{\edge}{\id{edge}}

\define{\grandparent}{\id{grandparent}}
\define{\parent}{\id{parent}}
\define{\father}{\id{father}}
\define{\mother}{\id{mother}}
\define{\personA}{\id{ann}}
\define{\personB}{\id{betty}}
\define{\personC}{\id{chris}}
\define{\personD}{\id{david}}

\usepackage{etoolbox}
\makeatletter
\preto{\@verbatim}{\topsep=0pt \partopsep=0pt }
\makeatother
\makeatletter
\patchcmd{\@verbatim}
  {\verbatim@font}
  {\verbatim@font\small}
  {}{}
\makeatother

\begin{document}
\maketitle



\begin{abstract}
In this paper we describe an approach to programming
micro\-controllers based on the Arduino platform
using Datalog as a clear and concise description language for system behaviors.

The application areas of cheap and easily programmable microcontrollers, like robotics,
home automation, and IoT devices hold mainstream appeal
and are often used as motivation in natural science and technology teaching.
The choice of programming languages for microcontrollers is severely limited,
especially with regard to rule-based declarative languages.

We use an approach that is based on the Dedalus language
augmented with operations that allow for side-effects
and takes the limited resources of a microcontroller into account.

Our compiler and runtime environment allow to run Datalog programs on Arduino-based systems.
\end{abstract}


\section{Introduction}

Logic and declarative programming is often and successfully used as parts of desktop and server applications.
We value the declarative techniques because it is easier to write programs that relate closely to the specification (or even write compilable specifications) and show their correctness.
Declarative programming has found its place in most computer science curricula in some form (often Haskell and Prolog) as well.
But especially in logic programming the applications often are theoretical or only used as part of a larger system.
The parts that interact with the outside world are usually written in an imperative fashion.
For embedded systems, where rule-based interaction with the outside world is often the majority of the application, declarative programming is an avenue not well explored.

With the advent of really cheaply produced microchips that allow for direct hardware interaction,
small and easily programmable systems have found a place in STEM education
(Science, Technology, Engineering, Math)
and are used
to teach electrical engineering, signal processing, mechanical engineering, robotics, and of course, programming in all levels of school and academia~\cite{DBLP:conf/rie/AgatolioM16,DBLP:conf/teem/Martin-RamosSLS16,DBLP:conf/iticse/RussellJS16}.
Systems like this are ubiquitous in the hobbyist realm and are most often used in IoT (Internet of Things) devices and home automation.
We can categorize those systems and the manner in which they can be programmed the following way:

\begin{itemize}
\item
\textbf{System on a chip} (SoC) devices like the Raspberry~Pi can, in principle,
be programmed using any software a traditional desktop computer can be programmed with.
While the available resources of SoCs are limited, the available main memory
is in the dozens or hundreds of megabyte and even the slowest devices have
CPUs with at least 300~MHz while the faster ones use multicore architectures
with operating frequencies in the gigahertz range.
Those CPUs are often found in phones, tablets, TVs, and other multimedia devices as well.
And since the SoC devices usually also have a standard-compliant Linux distribution
installed, any programming interface suitable to work with the GPIO (General Purpose Input Output)
can be used for the described tasks.
There is hardly any mainstream programming language that can not be used to program a Raspberry~Pi
or similar SoC devices.
Even the LEGO Mindstorms EV3 platform falls into this category
and students can engage with this platform using (among others) Python, Java,
Go, C, Ruby, Perl, and even Prolog~\cite{Sw17}
as their programming language of choice.
Almost any technology stack
for declarative logic programming
can be used on these devices.
\item
In contrast the ways in which \textbf{microcontrollers} can be programmed is very limited.
Microcontrollers often use 8-bit CPUs with
operating frequencies range from 16 to 40~MHz and an operating memory of 0.5 to 8~KB.
Even the larger microcontrollers like the ESP-family with 32-bit CPUs, operating frequencies of up to 240~Mhz and 520~KB of memory are unsuitable for a modern Linux kernel and userspace,
let alone the technology stack for declarative programming.
For this kind of embedded programming there have traditionally been only two options.
The approachable method that is often used in teaching beginner and intermediate
courses is a graphical
block-based
programming language like scratch that uses an
approach of translating code templates that fit like puzzle-pieces to actual C source code.
The second approach that is taken on academic or advanced level is to program
C code directly.
Both approaches limit the user with regards to available programming paradigms.
Imperative programming seems to have no real alternatives,
even though such systems that can be equipped with sensors, buttons, lights, displays, etc.
are, in principle, well-suited to be programmed using other paradigms.
Especially in interactive applications like environmental sensing and robotics,
event-driven or rule-based declarative approaches are desirable.
\end{itemize}

There have been many attempts to bring declarative programming to embedded systems.
Some declarative approaches, like LUSTRE~\cite{Halbwachs91thesynchronous} from the early '90s,
aim at reactive and dataflow oriented programming.
Comparative experiments with implementations of embedded applications using
abstract declarative languages (Prolog, OCaml) showed that while
the abstract code is shorter, the overhead for the runtime environments
is significant~\cite{DBLP:conf/sbcci/SpechtRCLCW07}.

In the recent past there have been advances in bringing event-driven programming
in the form of functional reactive programming (FRP) to the Arduino platform.
The Juniper programming language~\cite{DBLP:conf/icfp/HelblingG16} is such a language
that leverages the functional reactive style. The \texttt{frp-arduino} project%
\footnote{\url{https://github.com/frp-arduino/frp-arduino}} provides a domain-specific language
that is embedded into Haskell in order to create and compile FRP programs for
the Arduino.


There are other declarative programming approaches for the Arduino-based microcontroller platform
like Microscheme\footnote{\url{https://github.com/ryansuchocki/microscheme}}, a Scheme subset for the Arduino platform.
In the home automation context there have been projects that allow to declaratively configure
microcontroller systems with common sensor setups
(like ESPHome\footnote{\url{https://esphome.io/}})
but this approach is limited to this specific domain and a small number of targeted devices and peripherals.

But in terms of logic programming the Arduino platform is sorely lacking.
Logic programming languages like Datalog allow concise and clear descriptions
of system behaviors.
To use rule-based systems in the domain of robotics and home automation is very appealing.

In this paper
we propose a specific dialect of Datalog closely related to the Dedalus language~\cite{DBLP:conf/datalog/AlvaroMCHMS10} (Section \ref{sec:dedalus})
that includes IO operations.
We define an evaluation order for the different types of rules (Section \ref{sec:evaluation}) and
give a scheme to compile the Datalog code to C~code (Section \ref{sec:compilation}).
This scheme can be used to program Arduino-based microcontrollers in an
expressive and declarative fashion
which we show by providing some example programs (Sections \ref{sec:exampleprograms} and \ref{sec:macros}).

\section{Target Platform}

As our target platform we have chosen microcontrollers with the ATmega328 8-bit processor,
like the Arduino Nano, Arduino UNO\footnote{\url{https://www.arduino.cc}}, or similar devices
(see Figure~\ref{fig:arduino_picture}).
The ATmega328 is comparatively cheap and widely used.
This target platform comes with a set of limitations and design challenges:

\begin{figure}
  \begin{center}
    \includegraphics[height=3.5cm]{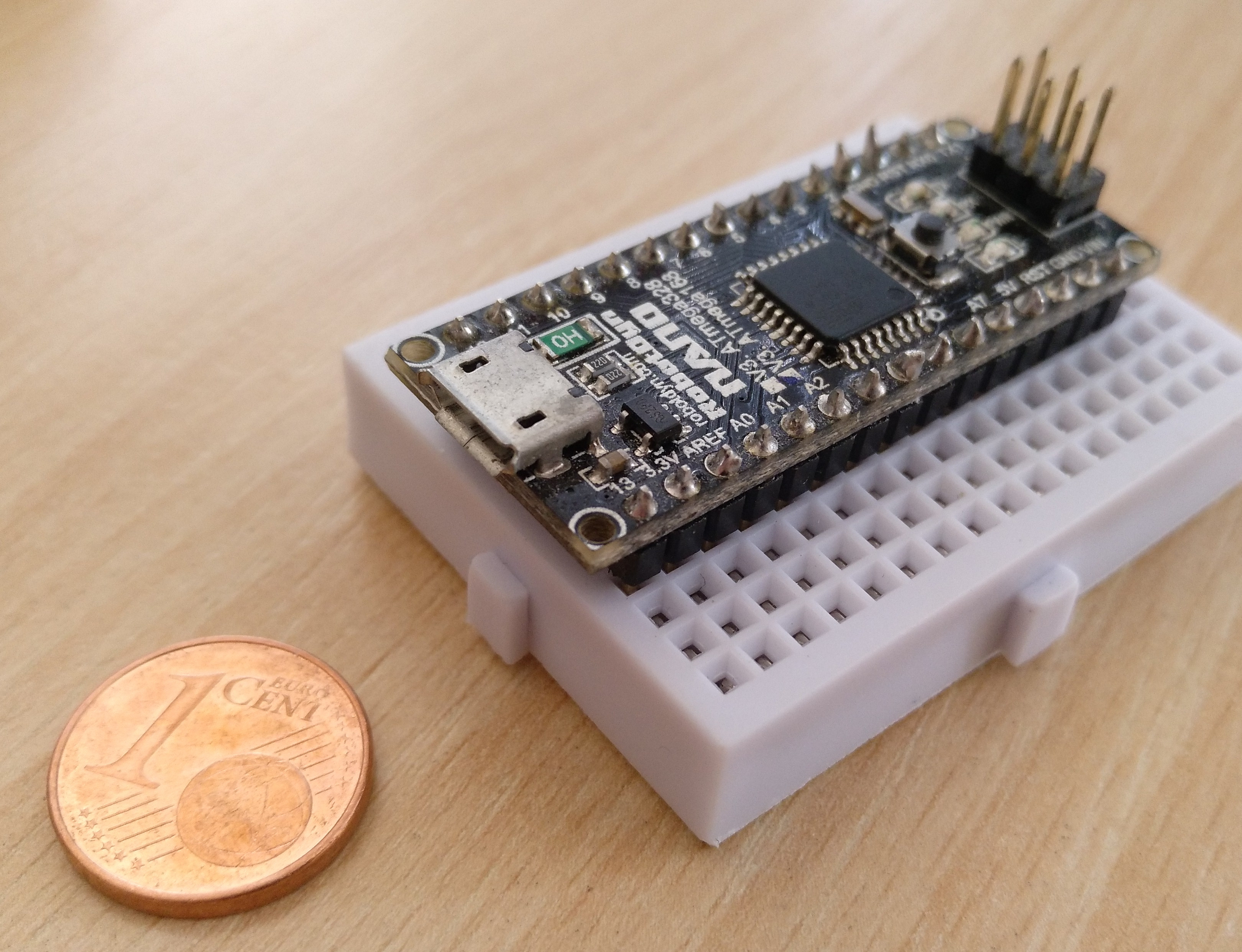}
    \includegraphics[height=3.5cm]{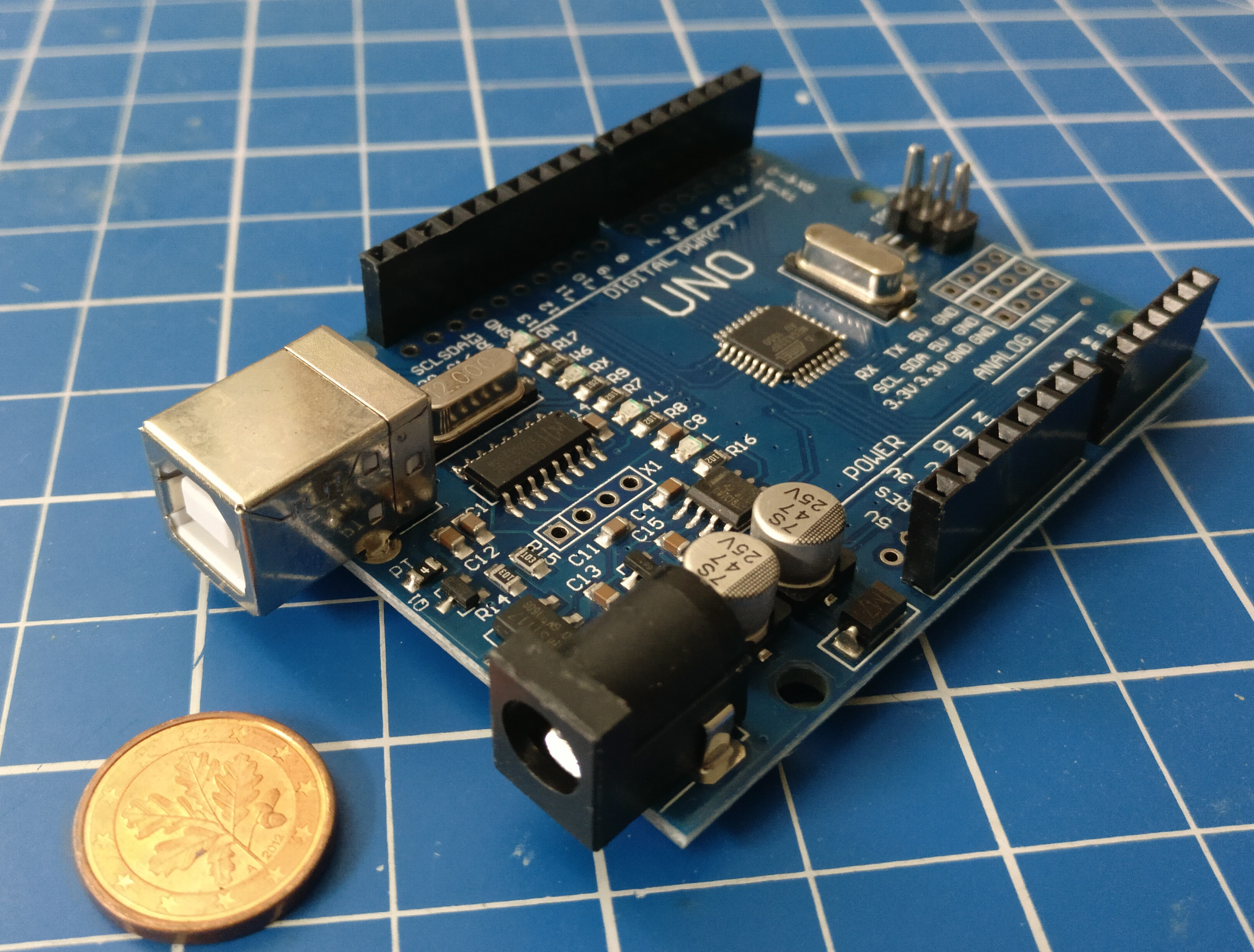}
  \end{center}
  \caption{Arduino Nano and Uno Compatible Boards with 1 Euro Cent for Size Comparison}
  \label{fig:arduino_picture}
\end{figure}

\begin{itemize}
  \item There is only 2~KB of SRAM available that is used for both heap and stack data.
        This means we are limited in operational memory for storing derived facts and
        in algorithm design with regards to function call depth.
  \item 32~KB of Flash memory can be used to store the program.
        This might seem a lot in comparison but this is also used to store
        additional libraries for peripheral access that are wanted by the user.
        This is also quite limiting considering the algorithm design
        and the amount of source code we are allowed to generate.
        The \texttt{Arduino.h} header files with pin input and output
        and writing to the serial port already use 2~KB of that memory,
        when compiled with size optimization enabled.
  \item A boot loader of about 2~KB is used for the firmware.
  \item The ATmega328 processor has an operational speed of 20~MHz
        which is a lot compared to the amount of data we have to operate on.
  \item There is an additional EEPROM non-volatile storage of 1~KB. 
        This storage is slow and is limited in the amount of write cycles.
        If the user chooses to write to or read from this storage as an effectful
        operation (i.e. IO~predicate, see Section~\ref{sec:dedalus}), they can do so.
\end{itemize}

The chosen target platform gives us restrictions with regards to the resource usage to aim at.
Since we generate C-code and our approach to interfacing with the rest of the system
is generic our approach works for other embedded systems and processors as well.
The generic approach is also useful since there already is a huge ecosystem for embedded development.
The ``PlatformIO'' platform\footnote{\url{https://platformio.org/}}
(self-proclaimed ``open source ecosystem for IoT development'')
has well over 600 different supported boards and over 6.400 libraries in its registry.
There is no reason why this effort should be duplicated.

\section{Extension to Dedalus language}
\label{sec:dedalus}

We base our work on the Dedalus\textsubscript{0} language (from here on just Dedalus).
Dedalus is a special variant of Datalog with negation
where the \textbf{final attribute of every predicate} is a ``timestamp''
from the domain of the whole numbers.
We call this attribute the ``time suffix''.
We give a quick overview over the Dedalus language~\cite{DBLP:conf/datalog/AlvaroMCHMS10}:

\begin{itemize}
  \item Every subgoal of a rule must use the same variable $\mathcal{T}$ as time suffix.
  \item Every rule head has the variable $\mathcal{S}$ as a time suffix.
  \item A rule is \textbf{deductive} if $\mathcal{S}$ is bound to $\mathcal{T}$,
        i.e. $\mathcal{S} = \mathcal{T}$ is a subgoal of this rule.
        \\ Example: $p(X, \mathcal{S}) \leftarrow q(X, Y, \mathcal{T}), p(Y, \mathcal{T}), \mathcal{S} = \mathcal{T}.$
        \\ We allow for stratified negation in the deductive rules.
  \item A rule is \textbf{inductive} is $\mathcal{S}$ is bound to the successor of $\mathcal{T}$,
        i.e. $successor(\mathcal{T}, \mathcal{S})$ is a subgoal of this rule.
        \\ Example: $p(X, \mathcal{S}) \leftarrow q(X, Y, \mathcal{T}), p(Y, \mathcal{T}), successor(\mathcal{T}, \mathcal{S}).$
        \\ We allow arbitrary negated body literals in inductive rules,
        because the program is always stratified with regards to the last component.
\end{itemize}

\noindent
In Dedalus every rule is either deductive or inductive.
To make it easier to work with those restrictions some syntactic sugar is added:

\begin{itemize}
  \item For deductive rules the time argument is left out in the head of the rule and every subgoal.
  \\ Example: $p(X) \leftarrow q(X, Y), p(Y).$
  \item For inductive rules the suffix ``@next'' is added to rule head
        and the time argument is left out in the head of the rule and every subgoal.
        \\ Example: $p(X)@next \leftarrow q(X, Y), p(Y).$
  \item For facts any timestamp of the domain is allowed as $\mathcal{S}$ (written using the @-notation).
        To keep the memory footprint low we only allow facts for the timestamp 0 in this notation.
        \\ Example: $p(5)@0.$
\end{itemize}

If a fact is not transported from one timestamp to the next we have a notion of deletion.
But Dedalus is more than just Datalog with updates.
With this extension our Datalog program now has a notion of time
where not everything happens at once
but the facts with some timestamp $T_n$
can be seen as ``happening earlier''
than the facts with timestamp $T_m$ with $n<m$.
Depending on the evaluation strategy, any fact with an earlier timestamp
may be deduced before those with a later timestamp.
The timestamp also captures a notion of state, similar to the Statelog language~\cite{DBLP:conf/dagstuhl/LausenLM98a}.
This is useful for interactions with the environment.

To facilitate this interaction we add a predicate type and two types of rules
that are used to manage effectful functions of the system (IO):

\begin{itemize}
  \item An \textbf{IO predicate} is a predicate that corresponds to a system function
        that has effectful behavior with regards to the environment.
        IO~predicates do not correspond to members of the minimal model of our program.
  \item An \textbf{IO~literal} is a literal from an IO~predicate.
        Depending on usage this can be considered a (restricted) variant of an action atom or external atom~\cite{DBLP:journals/ai/EiterSP99,DBLP:conf/ijcai/EiterIST05}.
  \item An \textbf{input rule} is a \textbf{inductive rule} that has as last subgoal a positive IO~literal
        corresponding to a system function that reads a value from the environment,
        like the current time or a sensor value.
        The system function is executed when it is needed to derive a fact for the next state.
  \item An \textbf{output rule} is a \textbf{deductive rule} that has an IO~literal as the head.
        The literal corresponds to a system function that changes the environment,
        like setting the output current of a pin.
        The system function is executed when the literal can be derived.
  \item A rule has at most one IO literal in either head or body.
  \item We also allow arithmetic comparison of bound variables and arbitrary arithmetic expressions within the operands of the comparison.
\end{itemize}

\section{Program Evaluation}
\label{sec:evaluation}
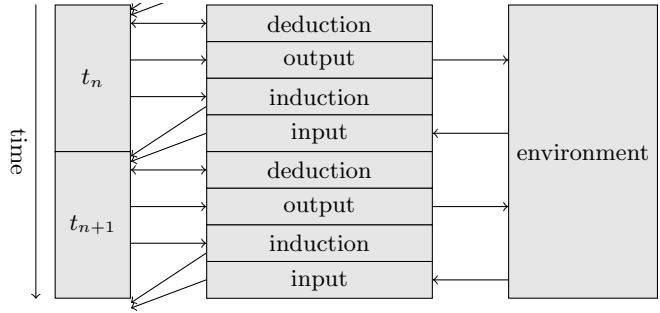
\begin{figure}
\tikzset{state/.style={draw, fill=black!10, minimum width=1cm, minimum height=6em}}
\tikzset{op/.style={draw, fill=black!10, minimum width=3cm, minimum height=1.5em, node distance=1.5em}}
\begin{center}
\begin{tikzpicture}[auto,]
  \clip (-2,1) rectangle + (10,-4.1);
  \node [state] (curr) {$t_n$};
  \node [state, below of = curr, node distance = 6em ] (next) {$t_{n+1}$};

  \node [op, right = 1cm of curr.north east, anchor=north west] (curr_deduction) {deduction};
  \node [op, below of = curr_deduction] (curr_output) {output};
  \node [op, below of = curr_output] (curr_induction) {induction};
  \node [op, below of = curr_induction] (curr_input) {input};
  \node [op, below of = curr_input] (next_deduction) {deduction};
  \node [op, below of = next_deduction] (next_output) {output};
  \node [op, below of = next_output] (next_induction) {induction};
  \node [op, below of = next_induction] (next_input) {input};

  \node [minimum height=12em, fill=black!10, draw, right = 1cm of curr_deduction.north east, anchor=north west] (env) {environment};

  \draw [->] (env.west|-curr_input) -> (curr_input.east);
  \draw [<->] (curr_deduction.west) -> (curr.east|-curr_deduction);
  \draw [<-] (env.west|-curr_output) -> (curr_output.east);
  \draw [<-] (curr_output.west) -> (curr.east|-curr_output);
  \draw [->] (curr.east|-curr_induction) -> (curr_induction.west);
  \draw [->] (curr_induction.-175) -> (next.61);
  \draw [->] (curr_input.west) -> (next.59);
  \draw [->] ($(curr_induction.-175) + (0,6em)$) -> ($(next.61) + (0,6em)$);
  \draw [->] ($(curr_induction.-175) - (0,6em)$) -> ($(next.61) - (0,6em)$);
  \draw [->] ($(curr_input.west) + (0,6em)$) -> ($(next.59) + (0,6em)$);
  \draw [->] ($(curr_input.west) - (0,6em)$) -> ($(next.59) - (0,6em)$);

  \draw [->] (env.west|-next_input) -> (next_input.east);
  \draw [<->] (next_deduction.west) -> (next.east|-next_deduction);
  \draw [<-] (env.west|-next_output) -> (next_output.east);
  \draw [<-] (next_output.west) -> (next.east|-next_output);
  \draw [->] (next.east|-next_induction) -> (next_induction.west);

  \draw [->] ($(curr.north west) - (0.25, 0)$) -- node[midway,sloped,below] {time}  ($(next.south west) - (0.25, 0)$);

\end{tikzpicture}
  \end{center}
  \caption{Fact Deduction Order}
  \label{fig:deductionOrder}
\end{figure}

\noindent
Deduction of facts for the state $t_n$, the following state $t_{n+1}$, and scheduling and execution of effectful functions happens
in 4 phases (see Figure~\ref{fig:deductionOrder}):

\begin{enumerate}
  \item In the deduction phase all facts for the current timestamp are derived.
        During this phase only the deductive rules (i.e. the rules that derive facts for the current timestamp) are used.
        In our case we use a naive evaluation strategy (taking the strata into account)
        that uses the least amount of additional memory but any datalog evaluation strategy that computes the fixpoint can be used to derive the facts for the current timestamp.
  \item In the output phase IO~functions that write data or affect the environment can be executed.
        Output rules of the form $B \leftarrow A_1 \wedge A_2 \wedge \dots \wedge A_n.$ where
        $B$ is the single IO~literal are evaluated in the order they are written
        in the Datalog program.
        The function corresponding to $B\theta$ is evaluated
        once for every ground substitution $\theta$
        for $A_1$ to $A_{n}$ where $A_1\theta \dots A_{n}\theta$ is in the minimal model.
  \item In the induction phase all facts for the next timestamp are derived.
        During this phase only the inductive rules (i.e. the rules that derive facts for the next timestamp) are used.
        Since facts derived through inductive rules may only depend on facts from the current timestamp,
        all necessary facts are known after one execution of each rule.
        Therefore all inductive rules are evaluated once (and in any order) for this timestamp.
  \item In the input phase IO~functions that read data from the environment can
        be executed.
        Input rules of the form $B \leftarrow A_1 \wedge A_2 \wedge \dots \wedge A_n.$ where
        $A_n$ is the single IO~literal are evaluated in the order they are written
        in the Datalog program.
        The function corresponding to $A_n\theta$ is evaluated 
        once for every ground substitution $\theta$
        for $A_1$ to $A_{n-1}$ where $A_1\theta \dots A_{n-1}\theta$ is in the minimal model
        of the current state and the derived $B\theta$ is in the minimal model for the following state.
\end{enumerate}

The effectful functions corresponding to IO~literals are called once for every
ground substitution of the free variables in the rest of the rule.
This is not a restriction since deduplication can be made explicit by introducing
additional rules.
Let $I$ be the IO~literal in the output rule $I \leftarrow p(X)$
then the function corresponding to $I$ is executed for every $X$ with $p(X)$ in the minimal model, even though $X$ does not appear in $I$.
With the introduction of the regular predicate $I'$ this can be rewritten to remove multiple execution as the rules $I' \leftarrow p(X)$ and $I \leftarrow I'$ which has only ever one or no ground substitution.
Then the needed memory for the duplicate checks is explicit and transparent for the programmer.

Since order is enforced in the input and output phase of the program we deviate from a purely declarative description.
It is not often the case that the order of gathering data from the environment or changing pin states is important.
In case it is, the order is explicit to the programmer.

Note that while we allow arithmetic comparison with arbitrary arithmetic expressions, new constants are only
introduced by input rules.
Since the number of facts for a specific timestamp generated by input rules is limited, the number of new constants introduced is finite as well.
While termination does not hold for the whole program
(and we do not want it to), the minimal model for
any specific state is always finite.
We say that our program is locally terminating, meaning that every following
timestamp is reached eventually.

\section{IO~Literals and Example Programs}
\label{sec:exampleprograms}

Our application is statically typed and we only allow primitive types
for our data values.
This is why all predicates need to be declared beforehand with
the static types of their arguments.
As syntax we use something similar to what the Souffl\'e system~\cite{DBLP:conf/cav/JordanSS16}
does to declare
relations.
\texttt{.decl r(unsigned long, byte)} declares the predicate \texttt{r} with
two arguments and their respective types.
Since we have no general mechanism for textual output of relations,
we do not need to define names for the arguments.

Before we can show example programs we want to give some IO~predicates for the Arduino interface.
Users can write their own IO~predicates to interface with any number of existing libraries for their system.
On a most basic level, an embedded board communicates with the outside world by means of GPIO-pins
(general purpose input/output) that are attached to sensors, actors, or other mechanical or electrical components.
Basic interface functions\footnote{\url{https://github.com/arduino/ArduinoCore-avr/blob/master/cores/arduino/Arduino.h}} for the pins in an Arduino-based systems (see Figure \ref{fig:pininterface}) are the functions
\texttt{pinMode} that sets whether a pin is in input or output mode,
\texttt{digitalWrite} that sets the output voltage (usually between the constants \texttt{HIGH} and \texttt{LOW}) of a pin
(both persistent until the next call),
and
\texttt{digitalRead} that reads the voltage on a pin and gives either a \texttt{LOW} or \texttt{HIGH} value.
\begin{figure}%
  \begin{verbatim}void pinMode(uint8_t pin, uint8_t mode);
void digitalWrite(uint8_t pin, uint8_t val);
int digitalRead(uint8_t pin);
unsigned long millis(void);\end{verbatim}%
  \caption{Extract from \texttt{Arduino.h} Header Files}
  \label{fig:pininterface}
\end{figure}%

We define an IO predicate, which always starts with an \texttt{\#} to denote that it is an IO~predicate,
with its arguments (left side)
by arbitrary C statements (right side).
Every IO~predicate may only have one definition.
Within the defining C-statements variables from the predicate arguments can be used (prepended with \texttt{\#} as to not overlap with constants like \texttt{HIGH} and \texttt{LOW}).
Constants from the outside C-code may also be used as constants in the Datalog-code using \texttt{\#} as a prefix.
These base functions are part of our standard library but since there are many different community-created libraries, we allow arbitrary C-code for interaction with our Datalog system.

\begin{figure}
\begin{verbatim}
#pinIn(P)  = {pinMode(#P, INPUT);}
#pinOut(P) = {pinMode(#P, OUTPUT);}
#digitalWrite(P, Val) = {digitalWrite(#P, #Val);}
#digitalRead(P, Val) = {int Val = digitalRead(#P);}
#millis(T) = {unsigned long T = millis();}
\end{verbatim}
\caption{Defined IO Predicates from the Standard Library}
\end{figure}
%

Depending on the definition of an IO~predicate, some binding patterns for IO~literals are not allowed.
Consider the IO~predicate defined as
\begin{center}\texttt{\#digitalRead(P, Val) = \{int Val = digitalRead(\#P);\}}.\end{center}
We say that the variable \texttt{P} is read in the definition (as its value is used in a function call) and the variable
\texttt{Val} is set in the definition.
This corresponds to binding pattern bound-free.
Every variable that is not read in the definition is considered set in the definition.
Some restriction arise from inserting the definitions into the source code ``as is'':

\label{sec:iorules}
\begin{itemize}
  \item If the IO~literal is used as the head of a rule, all variables appearing must be bound and read in the definition.
        The definition is compiled ``as is''.
  \item If the IO~literal is used in an input rule, all variables read in the definition must be bound by the other literals in the query.
        If some variable is set in the definition and bound by other literals,
        we compile the use of $p(A)$ with $A$ bound but set in the definition as $p(A'), A'=A$.
        When the rule is rewritten this way, the variable set in the definition is free again and we use a later comparison to check whether the values are equal.
\end{itemize}

\begin{figure}
\begin{minipage}[t]{0.4\textwidth}
\begin{verbatim}
% Predicates
.decl setup
.decl pressed

% Setup and Initialization
setup@0.
#pinIn(2) :- setup.
#pinOut(13) :- setup.
\end{verbatim}
\end{minipage}
\begin{minipage}[t]{0.6\textwidth}
\begin{verbatim}
% Input
pressed@next :- #digitalRead(2, #HIGH).


% Output
#digitalWrite(13, #HIGH) :- pressed.
#digitalWrite(13, #LOW) :- !pressed.
\end{verbatim}
\end{minipage}
\caption{Program That Changes the Led When a Button Is Pressed}
\label{fig:touchblink}
\end{figure}
\noindent
We give an example program that switches an LED
(the internal LED on this example board is connected to pin 13)
on when the button connected to pin 2 is pressed,
and off when it is released (see Figure \ref{fig:touchblink}).
This program only has input and output rules and defines a minimal behavior
that can easily be adapted to arbitrary connected sensors (temperature, distance) and
actors (relays, motors).

In the same manner we define the blink-program (see Figure \ref{fig:blinklarge}) that toggles the LED every second
using the system function \texttt{millis} that returns the number of milliseconds
since the microcontroller has been turned on.
In this example we use the deduction phase to deduce actions and depending on those
actions we both affect the environment and change the following state.
The switching action (\texttt{turn\_on} and \texttt{turn\_off}) is deduced explicitly and
the current state (\texttt{on\_since} and \texttt{off\_since}) is passed into the following state through the
inductive rules until the decision to toggle is reached.

\begin{figure}
\begin{minipage}[t]{0.4\textwidth}
\begin{verbatim}
% Declarations
.decl setup
.decl now(unsigned long)
.decl off_since(unsigned long)
.decl on_since(unsigned long)
.decl turn_off
.decl turn_on

% Setup and Initialization
setup@0.
#pinOut(13) :- setup.
off_since(0)@0.
now(0)@0.
\end{verbatim}
\end{minipage}
\begin{minipage}[t]{0.6\textwidth}
\begin{verbatim}
% Deduction
turn_off :- on_since(P), now(T), P+1000 < T.
turn_on :- off_since(P), now(T), P+1000 < T.
% Induction
on_since(P)@next :- !turn_off, on_since(P).
on_since(T)@next :- turn_on, now(T).
off_since(P)@next :- !turn_on, off_since(P).
off_since(T)@next :- turn_off, now(T).
% Input
now(T)@next :- #millis(T).
% Output
#digitalWrite(13, #HIGH) :- turn_on.
#digitalWrite(13, #LOW) :- turn_off.
\end{verbatim}
\end{minipage}
\caption{Blink-Program}
\label{fig:blinklarge}
\end{figure}

\section{Macro Expansion}
\label{sec:macros}

In the context of home automation and IoT some tasks are quite common
and need to be accomplished in many projects.
Some of these tasks are initialization of sensors,
persisting of facts into the EEPROM, or
delayed deduction of facts, like one second in the future.
To facilitate this, we allow for macro expansion in our programming language.
Macros are written in square brackets and are placed in front of a rule.
The rule is then rewritten on a syntactic level to accomplish the task.
We give two macros as an example:

\begin{itemize}
  \item The setup-macro rewrites a rule \texttt{[setup]head.} to
        \texttt{head :- setup.} and adds the fact \texttt{setup@0}
        that marks the first state $T_0$ with the fact \texttt{setup}.
        This can be used for initialization of pins and sensors as well
        as initial state.
  \item The \texttt{[delay:1000]} macro (with any integer number) adds rules
        that deduces the fact in the future (as many milliseconds in the future).
        The rule \texttt{[delay:X]head(Args) :- body(Args)} is replaced by the following rules:
        \begin{itemize}
          \item Initial time fact: \texttt{now(0)@0.}
          \item Reading current time: \texttt{now(T)@next :- \#millis(T).}
          \item Deriving the fact that is to be delayed: \\ \texttt{delayed\_head(Args, Curr) :- body(Args), now(Curr).}
          \item Deriving the delayed fact when the delay time is reached: \\
                  \texttt{head(Args) :- delayed\_head(Args, Await),\\
                  \hphantom{head(Args) :-} now(Curr), Await+X <= Curr.}
          \item The rule that transports delay forward if the time is not yet reached:\\
          \texttt{delayed\_head(Args, Await)@next :- delayed\_head(Args, Await),\\
          \hphantom{delayed\_head(Args, Await)@next :- }now(Curr), Await+X > Curr.}
          \item New predicate declarations: ~~~~~~~ \texttt{.decl now(unsigned long)} \\
          \texttt{.decl delayed\_head(<former arguments>, unsigned long)}
        \end{itemize}
\end{itemize}

\noindent
These macros decrease the program size and using the presented macros we
can now show the final and very concise version (without the declarations) of our blink-program.
Note that the expanded version is slightly different to the hand-crafted version
of the blink-program but behaves the same.
\begin{figure}
\begin{minipage}[t]{0.5\textwidth}
\begin{verbatim}
[setup]#pinOut(13).
[delay:1000]turn_on :- turn_off.
#digitalWrite(13, #HIGH) :- turn_on.
\end{verbatim}
\end{minipage}
\begin{minipage}[t]{0.5\textwidth}
\begin{verbatim}
[setup]turn_off.
[delay:1000]turn_off :- turn_on.
#digitalWrite(13, #LOW) :- turn_off.
\end{verbatim}
\end{minipage}
\caption{Concise Blink-Program using Macros}
\end{figure}

\section{Runtime Environment and Compilation}
\label{sec:compilation}


\subsection{Memory Management}

Our runtime environment uses two buffers to store deduced facts.
One buffer is for the facts in the current state and the other
buffer is for the facts in the following state.
Since we discard facts from previous states and do not dynamically allocate memory,
this scheme allows us to
not store timestamp data or superfluous pointers at all.
For the state transition the buffers are switched and the buffer for the
next state is zeroed.
The buffer size is given by the user during compilation
as some unknown amount of memory might be needed for the other libraries and their data structures.
Our facts are stored in the buffers in a simple manner:

\begin{itemize}
  \item Predicates are numbered (from 1) and we use the first byte to store the predicate (up to 255 different predicates).
  \item Subsequent bytes are used for the arguments.
  \item Facts are stored one after the other in the buffer.
  \item Empty tail of the buffers are filled with zeroes.
\end{itemize}

\begin{figure}
\tikzset{field/.style={draw, fill=black!10, minimum width=1.5em, minimum height=1.5em}}
\tikzset{brace/.style={decoration={brace},decorate}}
\tikzset{bnode/.style={midway,above}}
\tikzset{mbnode/.style={bnode,below,yshift=-0.05cm}}
\tikzset{mbrace/.style={brace,decoration={mirror}}}
\begin{center}

\begin{tikzpicture}[auto,node distance=0]

  \node [field] (p1_p) {$001$};
  \node [field, right = of p1_p] (p1_1) {$003$};
  \node [field, right = of p1_1] (p1_2) {$232$};
  \node [field, right = of p1_2] (q1_p) {$002$};
  \node [field, right = of q1_p] (q1_1) {$042$};
  \node [field, right = of q1_1] (q1_2) {$000$};
  \node [field, right = of q1_2] (q1_3) {$012$};
  \node [field, right = of q1_3] (p2_p) {$001$};
  \node [field, right = of p2_p] (p2_1) {$128$};
  \node [field, right = of p2_1] (p2_2) {$012$};
  \node [field, right = of p2_2] (f1) {$000$};
  \node [field, right = of f1] (f2) {$000$};
  \node [field, right = of f2] (fr) {$\dots$};

  \draw [brace] (p1_p.north west) -- (p1_2.north east) node [bnode] {$p(1000)$};
  \draw [brace] (q1_p.north west) -- (q1_3.north east) node [bnode] {$q(42,12)$};
  \draw [brace] (p2_p.north west) -- (p2_2.north east) node [bnode] {$p(-12)$};
  \draw [brace] (f1.north west) -- (fr.north east) node [bnode] {free memory};

  \draw [mbrace] (p1_p.south west) -- (p1_p.south east) node [mbnode] {$p$};
  \draw [mbrace] (p1_1.south west) -- (p1_2.south east) node [mbnode] {$1000$};
  \draw [mbrace] (p2_1.south west) -- (p2_2.south east) node [mbnode] {$-12$};
  \draw [mbrace] (q1_1.south west) -- (q1_1.south east) node [mbnode] {$42$};
  \draw [mbrace] (q1_2.south west) -- (q1_3.south east) node [mbnode] {$12$};
  \draw [mbrace] (q1_p.south west) -- (q1_p.south east) node [mbnode] {$q$};
  \draw [mbrace] (p2_p.south west) -- (p2_p.south east) node [mbnode] {$p$};

\end{tikzpicture}
  \caption{Mapping Example with Declarations \texttt{.decl p(int)}, \texttt{.decl q(byte, int)}}
  \label{fig:memoryMapping}
  \end{center}
\end{figure}

\noindent
This memory management scheme is very simple but uses no additional memory on
pointers for organizing the data structure.
Fact access time is linear in the number of stored facts.
This is a reasonable compromise since we can not store many facts anyways.
Consider predicates with lengths of 8 Bytes.
If we want to use 800 Bytes of our RAM for fact storage we would allocate
400 Bytes per Buffer with 50 facts until the buffer is full.
Saving memory on facts, pointers, and call stack
by not using more complex data structures
is reasonable.

\subsection{Target Code}

The following functions we compile to C-code:

\begin{itemize}
  \item Switching buffers and clearing of a buffer.
  \item Writing values of the available data types to a buffer.
  \item Reading values of the available data types from a buffer.
  \item Inserting a fact into a buffer.
  \item Retrieving a fact position from a buffer according to the used binding patterns
        with the first argument for the start of the memory area to search in and one additional argument for every bound value.
        At least the pattern where every value
        is bound is used since we use it for duplicate checking on insert of facts.
        These functions return 0 if there is no fitting fact in the buffer.
  \item Reading an argument value from a fact given the fact position in a buffer.
\end{itemize}

\noindent
Additionally we compile the size and memory locations of the buffers (\texttt{curr\_buff}, \texttt{next\_buff}), the size of the facts depending on the predicate, and
the mapping from predicates to numbers as constants into the code,
effectively storing them in the program memory.

\begin{itemize}
  \item For every rule we generate a function without parameters that returns whether facts have been inserted.
  \item The generated function contains a nested-loop-join for every literal in the body
        with variables bound in order of appearance in the rule.
  \item The generated function contains an if-statement for every arithmetic comparison.
  \item Additionally we generate a duplicate check for the fact that is to be inserted,
        and an insertion statement.
\end{itemize}

\noindent
%
The code that we compile the rule $p(A) \leftarrow q(A),p(B), A < B$ to,
where all arguments are integers, is shown in Figure~\ref{fig:compiled_rule}.

\begin{figure}
\begin{tt}
\begin{tabular}{@{}|c|l@{}}
\multicolumn{1}{l}{}&\textcolor{Black}{%
	bool deductive{\U}rule{\U}1() {\LB}}\\
\multicolumn{1}{l}{}&\textcolor{Black}{%
	~~bool inserted{\U}facts = false;}\\
\multicolumn{2}{@{}l@{}}{}\\[-9pt]
\cline{1-1}
$q(A)$&\textcolor{MidnightBlue}{%
	~~size{\U}t~q1~=~curr{\U}buff;}\\
&\textcolor{MidnightBlue}{%
	~~while~((q1~=~q{\U}f(q1))~!=~0)~{\LB}~~~~~~~//~find~next~q-fact}\\
&\textcolor{MidnightBlue}{%
	~~~~int~A~=~q{\U}arg1(q1);~~~~~~~~~~~~~~~//~read~first~argument}\\
\multicolumn{2}{@{}l@{}}{}\\[-9pt]
\cline{1-1}
$p(B)$&\textcolor{Fuchsia}{%
	~~~~size{\U}t~p1~=~curr{\U}buff;}\\
&\textcolor{Fuchsia}{%
	~~~~while~((p1~=~p{\U}f(p1))~!=~0)~{\LB}~~~~~//~find~next~p-fact}\\
&\textcolor{Fuchsia}{%
	~~~~~~int~B~=~p{\U}arg1(p1);~~~~~~~~~~~~~//~read~first~argument}\\
\multicolumn{2}{@{}l@{}}{}\\[-9pt]
\cline{1-1}
$A\!<\!B$&\textcolor{OrangeRed}{%
	~~~~~~if~(A~{\LT}~B)~{\LB}}\\
\multicolumn{2}{@{}l@{}}{}\\[-9pt]
\cline{1-1}
$p(A)$&\textcolor{OliveGreen}{%
	~~~~~~~~if~(p{\U}b(curr{\U}buff,~A)~==~0)~{\LB}~//~duplicate~check}\\
&\textcolor{OliveGreen}{%
	~~~~~~~~~~insert{\U}p(curr{\U}buff,~A);~~~~~//~insertion}\\
\multicolumn{2}{@{}l@{}}{}\\[-9pt]
\multicolumn{1}{l}{}&\textcolor{Black}{%
	~~~~~~~~~~inserted{\U}facts~=~true;}\\
\multicolumn{2}{@{}l@{}}{}\\[-9pt]
$p(A)$&\textcolor{OliveGreen}{%
	~~~~~~~~{\RB}}\\
\cline{1-1}
\multicolumn{2}{@{}l@{}}{}\\[-9pt]
$A\!<\!B$&\textcolor{OrangeRed}{%
	~~~~~~{\RB}}\\
\cline{1-1}
\multicolumn{2}{@{}l@{}}{}\\[-9pt]
$p(B)$&\textcolor{Fuchsia}{%
	~~~~~~p1~+=~size{\U}of{\U}p;~//~advance~pointer~past~seen~fact}\\
&\textcolor{Fuchsia}{%
	~~~~{\RB}}\\
\cline{1-1}
\multicolumn{2}{@{}l@{}}{}\\[-9pt]
$q(A)$&\textcolor{MidnightBlue}{%
	~~~~q1~+=~size{\U}of{\U}q;~~~//~advance~pointer~past~seen~fact}\\
&\textcolor{MidnightBlue}{%
	~~{\RB}}\\
\cline{1-1}
\multicolumn{2}{@{}l@{}}{}\\[-9pt]
\multicolumn{1}{l}{}&\textcolor{Black}{%
	~~return~inserted{\U}facts;}\\
\multicolumn{1}{l}{}&\textcolor{Black}{%
	{\RB}}\\
\end{tabular}
\end{tt}
\caption{Compiled Rule $\textcolor{OliveGreen}{p(A)} \leftarrow \textcolor{MidnightBlue}{q(A)}, \textcolor{Fuchsia}{p(B)}, \textcolor{OrangeRed}{A < B}.$} %
\label{fig:compiled_rule}
\end{figure}

For inductive rules
instead of writing the fact to the buffer
corresponding to the current timestamp, it is written into the buffer
corresponding to the following (next) timestamp.
IO~literals are compiled ``as is'' according to the rules from Section~\ref{sec:iorules} with their usage replaced by the C-statements they are defined with.

\subsection{Compiled Source File}

The end result of the compilation process is a C source file that can
be compiled to machine code using the Arduino toolchain (for example PlatformIO
or the Arduino IDE\footnote{\url{https://www.arduino.cc/en/Main/Software}})
and has the following general format:

\begin{figure}
\begin{verbatim}
// includes
// Buffer Declarations
// Functions for Buffer Access
// Reading and Writing Facts
void setup() {
  // Buffer initialization
  // Facts for timestamp 0
}
void loop() {
  do { // deductive phase
    added_facts = false;
    added_facts |= deductive_rule_1();
    ...
    added_facts |= deductive_rule_i();
  } while (added_facts);
  output_rule_1(); inductive_rule_1(); input_rule_1();
  ...
  output_rule_j(); inductive_rule_n(); input_rule_m();
  switch_buffers();
}
\end{verbatim}
  \caption{Simplified Outline of Compiled Source File}
  \label{fig:compiled_outline}
\end{figure}

\noindent
The \texttt{setup} and \texttt{loop} functions are the entrypoints
for the processor. The setup-function is called once when the microcontroller is started
and the loop-function is called repeatably once the setup has finished.
The loop function executes all the derivation steps in the proper order (see Figure \ref{fig:compiled_outline}).

\section{Conclusion}

We have shown that programs for Arduino and similar microcontroller systems
can be written in a declarative logic language
with few restrictions
using a slightly altered version of the Datalog dialect Dedalus.
Effectful operations are introduced by defining an evaluation scheme
where local termination still holds.

The Dedalus approach seems useful as it not only captures a notion of state-changes
during the execution in an interactive environment,
the captured notion of time allows us to use IO functions depending on facts corresponding to the state we consider as ``now''.

Then we have presented a straightforward translation scheme
from our program code to Arduino-C that integrates well with existing
library functions.
While the generated code corresponds to a naive evaluation scheme,
it is not algorithmically complex and uses not too much of the available
program memory.

Additionally we showed a method for code expansion that extends the usefulness
of our language by autogenerating boilerplate code.
This means that introductory examples of Arduino programs
written in Datalog
are as easy,
if not easier than the equivalent C program.

There are still a few open questions and areas for further research.
Is it useful to apply transformations like magic sets,
SLDmagic~\cite{DBLP:conf/cl/Brass00}, or our Push method~\cite{DBLP:conf/ershov/BrassS17}
with the IO~rules as query goals?
How well do other datalog optimzation and compilation schemes work with the limited operating memory?

With a focus on the physical aspects of specific boards, can we analyze the
program to find pins that are used as input but defined as output and vice versa?
Can we identify otherwise incorrectly used system resources like pins that might
be set differently multiple times in the same state, or facts that may not
co-occur in the same state (like \texttt{led\_on} and \texttt{led\_off})?

The initial state of the program is known beforehand (there is no dynamic database for EDB facts)
and parameters are only introduced through input rules.
Can a set of possible states for the application, parameterized in the arguments of the facts,
be calculated beforehand and used as program state instead of a general purpose fact storage?

Since the memory on the chip is severely limited, can we give an upper
bound on the number of facts deduced for every timestamp
(e.g. the amount of memory needed for the runtime system)
using functional dependency analysis for derived predicates~\cite{DBLP:conf/lopstr/EngelsBB17}?
If this was known during compilation, the buffers can be appropriately sized automatically.
How quickly can the minimal model for a state be deduced and can we give
upper and lower bounds for the duration of one timestamp?
The last two questions are especially interesting with regards to real-time
applications and safety and liveness properties of embedded systems.

\bibliographystyle{splncs03}

\bibliography{wlp19}


\end{document}